\documentclass[10pt]{article}
\usepackage{ijcai17}
\usepackage{epsfig} 
\usepackage{amsmath}
\usepackage[hidelinks]{hyperref}
\PassOptionsToPackage{hyphens}{url}\usepackage{hyperref}
\usepackage[outdir=./]{epstopdf}
\usepackage{graphicx}
\title{Ranking ideas for diversity and quality}

\author{Faez Ahmed\\ 
Dept. of Mechanical Engg.\\
University of Maryland\\
faez00@umd.edu
\And
Mark Fuge\\ 
Dept. of Mechanical Engg.\\
University of Maryland\\
fuge@umd.edu
}

\newcommand{\eg}{{\em e.g.}}
\newcommand{\ie}{{\em i.e.}}
\newcommand{\etc}{{\em etc.}}
\newcommand{\etal}{{\em et~al.}}
\newcommand{\cutequationup}{\vspace*{+0.12in}}
\newcommand{\cutparagraphup}{\vspace*{+0.1in}}

\DeclareMathOperator*{\argmin}{arg\,min}

\begin{document} 

\maketitle    
\begin{abstract}
{\it 
When selecting ideas or trying to find inspiration, designers often must sift through hundreds or thousands of ideas.
This paper provides an algorithm to rank design ideas such that the ranked list simultaneously maximizes the quality and diversity of recommended designs.
To do so, we first define and compare two diversity measures using Determinantal Point Processes (DPP) and additive sub-modular functions. We show that DPPs are more suitable for items expressed as text and that a greedy algorithm diversifies rankings with both theoretical guarantees and empirical performance on what is otherwise an NP-Hard problem.
To produce such rankings, this paper contributes a novel way to extend quality and diversity metrics from sets to permutations of ranked lists.

These rank metrics open up the use of multi-objective optimization to describe trade-offs between diversity and quality in ranked lists. We use such trade-off fronts to help designers select rankings using indifference curves. However, we also show that rankings on trade-off front share a number of top-ranked items; this means reviewing items (for a given depth like the top 10) from across the entire diversity-to-quality front incurs only a marginal increase in the number of designs considered. While the proposed techniques are general purpose enough to be used across domains, we demonstrate concrete performance on selecting items in an online design community (OpenIDEO), where our approach reduces the time required to review diverse, high-quality ideas from around 25 hours to 90 minutes. This makes evaluation of crowd-generated ideas tractable for a single designer. Our code is publicly accessible for further research.
}
\end{abstract}

\section{Introduction}

When generating creative designs, both practicing designers and researchers agree: ``If you want to have good ideas, you must have many ideas." \cite{pauling2001linus} Why? Because having many ideas helps a designer\textemdash or a team of designers\textemdash explore a design space and find new inspiration from unlikely places.
But is more always better? When do `many ideas' turn into `too many ideas'?
Given thousands of possible ideas to process and limited time, a designer needs a much smaller ``good'' set of seed ideas or, better yet, a good ranking of all ideas so that they can decide when they have had enough.

But what, specifically, does it mean for a ranking of ideas to be ``good'' and how does one compute such rankings? This paper focuses on those two questions. Specifically, the paper argues that when ranking ideas\textemdash \eg, for the purpose of inspiration, idea generation, or selection\textemdash a good ranking should not only show a designer ideas that possess high \textit{quality}\textemdash that is, ideas that perform better than other ideas (assuming one can measure such differences accurately)\textemdash but also that possess \textit{diversity}\textemdash that is, a designer should see ideas that \textit{cover} a design space well.

Why would one care about encouraging diversity when ranking ideas? Why not just order ideas by individual quality or merit and be done with it?
Consider the following example design problem from a real-world design competition\footnote{{http://challenges.openideo.com/challenge/localfood/}} which asked designers to generate ideas to address ``How might we better connect food production and consumption?'' Of 606 submitted ideas, let us take a summary of just four ideas as an example:

\begin{enumerate}
\item Compost It!---A proposal to partner with the city to create a closed loop composting system.\footnote{{challenges.openideo.com/challenge/localfood/concepting/compost-it}} 
\item Residential compost material -- curbside pickup--- A state-wide initiative to encourage people to separate compost material for pick up.\footnote{{challenges.openideo.com/challenge/localfood/concepting/residential-compost-material-curbside-pickup}} 
\item The Art of Food Festival--- A festival celebrating local food and art with edible sculptures, inspired by french festivals. \footnote{{challenges.openideo.com/challenge/localfood/concepting/the-art-of-food-festival}}
\item Online local farming NFP organisation---Growing and delivering fresh locally grown vegetables to a community of online customers at a very low cost. \footnote{{challenges.openideo.com/challenge/localfood/concepting/online-local-farming-nfp-organisation}}
\end{enumerate}

The above ideas have quality scores\textemdash provided by human raters\textemdash of $20$, $12$, $9$, and $3$ points respectively. Our task is to show two ``good'' ideas to a designer where a ``good'' set of ideas should help inspire the designer to come up with new ideas.
One naive way is simply to order all ideas by their quality score and select the top two ideas.
However, in our example, this will select the two ideas related to composting. 
Is this a good choice?

On the one hand, they are the two highest quality ideas of the four.\footnote{Assuming (perhaps tenuously) that our measurement system, be it humans, computational simulations, analytical formulas, \etc~is not noisy, biased, or fixated towards particular solutions like composting.} On the other hand, they are surprisingly similar to each other; both address the fairly broad problem statement\textemdash connecting food production and consumption\textemdash via a narrow set of solutions\textemdash composting.
As many researchers have shown, generating good ideas requires both divergent and convergent thinking, and it is not clear that ranking purely by quality promotes such divergence. Likewise, if quality ratings are biased or noisy, promoting coverage may protect against unfairly discounting certain ideas. Ideally, selected ideas should have \textit{both} high quality and good coverage of possible options. This allows a designer to gain maximal benefit from a large number of ideas\textemdash \eg, increased coverage and quality\textemdash within a given budget of time or attention. 

How does one find high quality ideas that also have good coverage? One manual approach might first rank ideas by quality and then just swap ideas which are similar to each other with random ideas from the collection.
For our above example, the first two ideas are similar, so we can swap the second idea with either the third or fourth to get a diverse set of two ideas. 
But when the number of ideas grow to hundreds or thousands this approach does not scale; finding exactly which ideas to swap in is laborious and depends on the other ideas you already have in the set.
Astute readers may notice that, mathematically, this is equivalent to a combinatorial optimization problem called \textit{set covering} which is a type of boolean satisfiability problem. Optimizing such problems is NP-Hard.
Second approach, and one which is commonly used, is to define a objective function which is a weighed average of diversity and quality. While this approach is straightforward to implement, 
it is difficult to know beforehand how much quality one is willing to part with to encourage diversity.
Finally, the approach we use formulates a multi-objective optimization problem and treats coverage and quality as independent objectives.
One benefit of doing so is that after computing the trade-off front one can actually compute the loss in quality for any given gain in coverage.

In addition, as different designers may have different information needs, instead of selecting a smaller subset of two ideas and showing them to a designer, one can also \textit{rank order} all ideas. This retains all ideas where the ones appearing on top of the list are good (\ie, higher quality with good coverage). Deciding what ranking is better is non-trivial. Even for our simple example, it is hard to argue which of the following rankings is clearly better: [1,3,4,2] or [1,4,3,2] or [1,3,2,4]. While, at first glance, ranking ideas may seem straightforward, including diversity transforms ranking into an NP-Hard problem.

\paragraph{This paper's contributions}
We propose a practical, efficient, computational method for ranking diverse and high-quality items. 
In contrast with past work, we approach idea ranking as a multi-objective optimization problem, which allows a designer to trade off rankings between those that encourage diversity and those that encourage quality.
Specifically, the main research contributions of this paper are:
\begin{enumerate}
	\item We define a novel method for extending set-based diversity measures~\cite{ahmed:idetc_2016_idea} to rank-based diversity measures. Our key insight lies in how to preserve a mathematical property called \textit{sub-modularity} when computing diverse rankings; without it optimization becomes intractable.
    \item We propose a polynomial-time greedy algorithm to rank items by diversity. This algorithm has both theoretical approximation guarantees and outperforms existing benchmarks.
    \item We describe how to balance high-quality versus diverse idea rankings through a quality and diversity trade-off front among rankings.
    \item We evaluate two state-of-the-art approaches to compute diversity of item sets\textemdash sub-modular clustering and Determinantal Point Processes\textemdash uncovering the conditions under which one out-performs the other.  
\end{enumerate}

\paragraph{Structure of the paper}
We want a way to rank items that balances quality and diversity. While quality rankings are well-researched and comparatively tractable (see Sec.~\ref{sec:ranking_quality}), Diversity measures\textemdash typically defined over fixed-sized sets\textemdash are less straightforward. Before we can combine quality and diversity for ranking (Sec.~\ref{sec:ranking}), we need to first define diversity (Sec.~\ref{sec:defining_diversity}), including what it means to \textit{cover} a \textit{space} of ideas (\ref{sec:item_space}) and how to compute that coverage for a \textit{set} of ideas (\ref{sec:clustering} \& \ref{sec:dpp}). We then describe how to extend diversity and quality to \textit{rankings} (Sec.~\ref{sec:ranking}) rather than a fixed-size set. To compute such rankings, we introduce both global (Sec.~\ref{sec:global_opt}) and greedy (Sec.~\ref{sec:greedy}) optimizers that take advantage of properties of sub-modular functions to hasten convergence and provide theoretical performance guarantees. Section~\ref{sec:results} demonstrates our approach on real-world design ideas created by a crowd-sourced design community (OpenIDEO). Sec.~\ref{sec:discussion} adds discussion on main insights, addresses our key limitations and future work alongwith implications for design research, which include important choices in how we define similarity and quality as well as handling the multimedia nature of design ideas \ie, combinations of text, images, audio, \etc. The paper's supplemental material includes additional experiment that demonstrates applicability to ideas represented as sketches. It also includes an experiment which describe under what conditions one coverage metric outperforms another.

\section{Related work}
Two seemingly disparate fields\textemdash Design and Computer Science\textemdash have both explored ways to jointly rank quality and diversity. Design researchers have focused on appropriate metrics for measuring item diversity and quality, while Computer Science researchers have focused on representations and methods for scalably estimating or ranking lists of diverse items. This work advances different efforts across both fields.

Within Design, researchers have primarily tackled how to either (1) evaluate creative sets of ideas or (2) leverage large design databases to inspire designers. 
As an exemplar of the former, Shah \etal \cite{shah2000evaluation} provide metrics for ideation effectiveness, where the main measures for the goodness of a design method are how they expand the design space and how well they explore it. Typically, work in this vein discusses diverse design space exploration using terms like \textit{variety}, measured through, for example, coverage over trees of functions \cite{shah2000evaluation, verhaegen2013refinements}, human expert assessment \cite{cat}, or linear combinations of design attributes \cite{fuge2013automatically}. 
One of the difference between past engineering design variety literature and what we propose is that many past variety measures require expert coding for all ideas, which may be infeasible for a large collection.

The second main avenue of research concerns evaluating large sets of ideas, typically by using crowds of evaluators to scale up evaluation by partitioning ideas among many people. As an exemplar of such approaches, Kudrowitz and Wallace \cite{kudrowitz2013assessing} suggest metrics to narrow down a large collection of product ideas. Likewise, Green \etal \cite{green2014crowd} propose methods for creativity evaluation using crowd-sourcing, where researchers focused on inspiring designers \cite{von2005democratizing} and inspiring creativity \cite{chiu2012investigating}.

Within Computer Science, researchers have tackled diversification in two strongly inter-connected applications: information retrieval and recommender systems, where researchers have developed ranking algorithms for different settings. 
When recommending sets of items to people (\eg, movies on Netflix) predicting exactly what a user wants is difficult, so by recommending a diverse set of items, chances increase that one of the recommended items will match what the user wants. The intuition for this approach stems from the \textit{portfolio effect} \cite{ali2004tivo} where placing similar items together within a portfolio of items has decreasing additional value for users. This \textit{diminishing marginal utility} property is well-studied in consumer choice theory and related fields~\cite{coombs1977single}.

The main research questions within both recommender systems and information retrieval are two-fold: (1) how do we represent this diminishing marginal utility, and once we do (2) how do we optimize over it efficiently?
For the former question, researchers have proposed alternate scoring methods to diversify rankings. An early exemplar of this was Ziegler \etal \cite{ziegler2005improving} who modeled the topics in text documents and then tried to balance the topics within recommended lists. Their large scale user survey showed that a user's overall satisfaction with lists depended on both accuracy and the perceived diversity of list items. Approaches that followed largely centered around the notion of \textit{coverage}\textemdash that a diverse set should somehow cover a space of items well. The main differentiators of past approaches are how this coverage is measured and then combined with other objectives such as document relevance.

Approaches to measuring coverage break into two main camps, depending on what objects the coverage is defined over. The most common approach defines a vector space using properties of each item, \eg, word frequency vectors or topic distributions over text. For example, Puthiya \etal \cite{puthiya2016coverage} take positively rated items from a user, and then select sets from that list such that they cover the distribution of words in the submission. Likewise, search diversification techniques such as xQuAD \cite{santos2010exploiting} explicitly model the underlying aspects or subtopics for a query and select documents based on a combination of their relevance to the original query and relevance to the aspects. 

The second camp instead defines a similarity graph between items\textemdash for example cosine similarity between documents\textemdash and then computes properties over this graph such that the selected items maximize some graph coverage property. For example, one can use random-walk based algorithms like PageRank \cite{zhang2005improving,he2012gender} to estimate how central items are in a graph, and then re-order items based on this score. For more examples of such variants, Vargas \etal \cite{Vargas:2011:RRN:2043932.2043955} and Castells \cite{castells2015novelty} provide useful frameworks and reviews of past approaches. Such approaches apply to a broad set of applications like music discovery \cite{zhang2012auralist}, keyword-based summarization \cite{Fisher2015EvaluatingRD}, ecology\cite{patil1982diversity}, and document summarization \cite{zhu2007improving}.

Assuming we can answer the former question\textemdash how to represent diminishing marginal utility of sets\textemdash the latter question concerns computing such rankings. Three difficult and inter-related problems have motivated past research: (1) there are different ways of computing \textit{coverage} over a space\textemdash under what conditions would we prefer one over the other? (2) Coverage over \textit{sets} of items is a combinatorial problem (optimizing set-cover is NP-Hard)\textemdash how can we guarantee certain performance in polynomial time? And (3) diverse rankings require some notion of optimal coverage across a ranking, which is harder than guaranteeing coverage over a single fixed-size set\textemdash how should we compare optimal coverage over such rankings?

For the first problem of which coverage metric to use, researchers have proposed many different options. However not much work has characterized and compared the differences between these options; this is one of our paper's contributions. For the second problem, most work has focused on using greedy approximations to the set coverage problem. This means most of these methods produce a list by progressively adding items to a set, with some fixed weighted trade-off between diversity and relevance \cite{zhao2016much}. While this efficiently produces diverse lists, it is difficult to compare or customize such lists when users have different preferences between diversity and quality. One of this paper's contributions is to provide, to our knowledge, the first approach to compare entire ranked lists between these two objectives and efficiently create rank orders that span the trade-off between diversity and quality (Sec.~\ref{sec:ranking}). 
For the third problem, past work typically considers rankings more diverse if they minimize some notion of redundancy. For example, whether ranked items occur in common elements in a hierarchy \cite{Wang:2016:ESR:2911451.2911497}, or how well rankings compare with human relevance judgments of sub-topics such as ERR-IA \cite{chapelle2011intent}, $\alpha$-nDCG \cite{clarke2008novelty}, and S-precision or S-recall \cite{carterette2009analysis}. These metrics are difficult to extend to cases where we do not have human-provided labels. One of this paper's contributions is to extend coverage metrics used for fixed-size sets to rankings, such that we can use those metrics to evaluate diversity of ranked lists (Secs.~\ref{sec:ranking_diversity} and \ref{sec:ranking_quality}).

Compared to information retrieval or recommender systems, where the number of sub-topics is frequently set in advance and users have a specific query they wish to answer, design ideas are often unstructured, come from a wide variety of sources, and a designer's goal is to gain inspiration from a wide range of sources. This makes generating diverse, high quality lists particularly important when providing ranked ideas to designers.
If successful, such techniques would have wide ranging consequences for crowd-sourced or large-scale ideation techniques by helping designers avoid premature convergence on a very limited set of ideas and helping people explore vast design spaces.

\section{Defining and Computing Diversity for Fixed-Size Sets}
\label{sec:defining_diversity}
Before we can address \textit{ranking} ideas by diversity, we first need to introduce how to quantitatively compute the diversity for simpler \textit{fixed-sized} sets of ideas. For example, when one needs to pick a diverse set of five ideas, but the exact order in which one picks them does not matter.

Consider the example from the beginning of the paper, where one needs to select two ideas out of four related to ``connecting food production to consumption.'' In that example, one can intuitively tell that selecting the first two ideas\textemdash both relating to composting strategies\textemdash seems less diverse than the first and third ideas\textemdash one on composting, and one on food festivals. Why does one conclude this? How can we make this intuition more precise? Can we quantitatively capture that intuition?

As with the related work summarized above, quantitatively measuring diversity essentially comes down to measuring how well a set of ideas \textit{covers} a \textit{space of options}. For our above example, one might look at the four ideas and mentally place them into ``buckets,'' placing the two composting ideas into the ``compost'' bucket, the food festival idea into an ``events'' bucket, and the online farming group into an ``online community'' bucket. Computing diversity\textemdash or how well a set covers a space of options\textemdash might then translate into calculating whether selected ideas come from different buckets.

Alternatively, one could imagine printing out the ideas, placing them on a table, and moving them around such that similar ideas were close to one another and different ideas were far away. Computing diversity might then involve calculating whether selected ideas came from different parts of the table, spanned a large area of the table, \etc~
While different mathematical representations of design spaces and how to quantify their coverage may lead to different definitions of diversity, the central idea remains the same.

The rest of this section first reviews how to represent the space of options\textemdash namely, via a similarity function between ideas. Then it presents two existing state-of-the-art methods to compute coverage over that space\textemdash one that uses clustering (\ie, buckets) via additive sub-modular functions and one that uses on continuous spaces via Determinantal Point Processes. Our supplemental material presents additional experiments that compare the conditions under which one diversity measure outperforms the other.

While we selected the below methods to demonstrate our ranking approach on a concrete, real-world example, it is important to note that this paper's main contributions\textemdash how to combine quality and diversity measures to efficiently compute ideas rankings\textemdash do not depend on those specific choices. As we describe in more detail below, our ranking approach (Sec.~\ref{sec:ranking}) applies to any choice of design space representation and diversity coverage measure, provided that they satisfy two mild technical conditions.\footnote{In brief, 1) the space must allow one to compute a positive-semidefinite similarity function between points in the space and 2) the diversity function must be \textit{sub-modular} (\ie, obey diminishing marginal utility).}

\subsection{Representing ideas and their similarity}
\label{sec:item_space}
Before we can compute coverage over a space, we need represent ideas such that we can compute similarity between them. This is generally done in one of two ways.

The first and most common way is to explicitly represent ideas within a Hilbert space\textemdash \ie, a space that permits inner products, such as Euclidean space\textemdash and then compute how similar ideas are by taking inner products between them in that space. For example, one can represent geometry or CAD objects using a vector of parameters from a parametric model or using latent semantic dimensions learned from the geometry~\cite{chen2016designs, yumer2015procedural, burnap2016improving}. For images or sketches, one can use image processing techniques like SIFT features or deep learning (\eg, Sketch-a-Net \cite{yu2015sketch}) to transform free hand sketches to a vector space. For ideas expressed through text one can use bag of words or latent vector space models, such as Latent Semantic Analysis~\cite{dong2005latent}. For mixed-media designs, such as combinations of sketches and text, one can even learn joint vector space models~\cite{pu2016variational}. Similarity is then computed through, for example, cosine, jaccard, or squared euclidean distances between those two vectors.

The second way, and the one we have demonstrated in supplement material is to compute similarity between ideas directly using either a \textit{kernel function}\textemdash a function that, given two ideas, computes the similarity between\textemdash or by having humans directly rate the similarity between ideas~\cite{tamuz2011adaptively}. 
The former is useful in design when one wants to compute diverse, high-quality rankings of structured objects\textemdash that is, designs expressed as graphs or hierarchies, such as Function Structures~\cite{qian1996function} or Function Decompositions~\cite{kirschman1998classifying,stone2000development} using Graph Kernels~\cite{vishwanathan2010graph}. The latter is useful when ideas are too difficult or complex to easily describe using a set of analytical functions, but one has human experts on-hand who can provide similarity judgments (\eg, idea A is closer to idea B than C, \etc)~\cite{tamuz2011adaptively}. Through asking human experts (or crowd-sourcing the task), one can compute a kind a ``Human Kernel'' that can provide sufficient information for our below ranking technique to use. To further demonstrate our method for sketches, we have shown an example in supplement material with five sketches and human ratings to compute the trade-off front.

This paper's main contribution\textemdash an efficient ranking algorithm for high quality and diverse ideas\textemdash is agnostic to the above choice of similarity function. However, a similarity function or matrix, whether chosen analytically or computed by humans, does need to satisfy one mild technical condition\textemdash it must be positive-semidefinite. In practice, most widely used methods of computing similarity between vectors, such as cosine, radial basis function, or hamming distances satisfy this condition. If one wants to use their own similarity function, this condition is also straightforward to verify.

For the rest of the paper, we will assume, without loss of generality, that we can compute a symmetric similarity matrix $L$ whose entries $L_{i,j}$ correspond to the similarity between ideas $i$ and $j$, where $L_{i,j}=1$ means that ideas $i$ and $j$ are identical and $L_{i,j}=0$ means that the ideas are completely dissimilar. 

The next two sections introduce two existing, competing, state-of-the-art methods\footnote{As measured with respect to success at a common benchmark task of automatic document summarization (\eg, at the Document Understanding Conference \cite{Lin2011class,lin2012learning}), which require selecting high quality non-redundant sentences to summarize a document.} for computing diversity with respect to a similarity kernel. Specifically, sub-modular clustering \cite{Lin2011class,lin2012learning} and Determinantal Point Processes (DPPs) \cite{kulesza2012determinantal}, which correspond, respectively, to thinking about coverage over discrete ``buckets'' versus volumes in continuous spaces. Our supplemental material provides additional experiments that characterize the conditions under which one outperforms the other; we found that DPPs were a more robust choice for different problems and we use them for our experimental results later in the paper. 

\subsection{Clustering-based Diversification}
\label{sec:clustering}

One way to think about covering a space of ideas is to think about ideas as falling into different categories, types, clusters, or ``buckets." Diversity might then entail promoting adding ideas to empty buckets and penalizing selecting ideas all from one bucket. That is, we wish to model diminishing marginal utility\textemdash that adding an idea to a bucket where one already has lots of ideas is not as valuable as adding a (similar quality) idea to an empty bucket.

This is the approach Lin \etal~\cite{Lin2011class,lin2012learning} use, where they show that many existing diversity methods are instances of a \textit{sub-modular function}. Sub-modular functions are similar to convex functions, but defined over sets rather than the real line. Such functions are designed to model diminishing marginal utility, which is exactly the mathematical property one needs to model diversity \cite{fuge2013automatically}. We propose a metric inspired by the diversity reward function used by Lin \etal \cite{Lin2011class} for multi-document summarization, which rewards diversity of a set of items as shown below:

\cutequationup
\begin{equation}
Div_1(S) = \sum_{k=1}^K\sqrt{\sum_{j\in S\cap P_k}\frac{1}{N\times M}\sum_{i\in P_k}L_{i,j}}
\label{eq_div}
\end{equation}

\cutparagraphup
Here, $V = {v_1, . . . , v_n}$ is the set of all $N$ items in a set. Subset $S \subseteq V = {s_1,...,s_m}$ is the selected $M$ items given $K$ clusters. $P_i$, i = 1,...,K is a partition of the ground set $V$ into separate clusters (\ie, $\cup _i P_i = V$ and the $P_i$s are disjoint). 
That is, an item can only belong to one cluster. 
The square root function automatically promotes diversity by rewarding items from clusters which have not yet contributed items.

To understand above metric, let us take our example, where the collection has three known topics\textemdash compost,
food festivals, and online web communities. For illustration purposes, consider that adding an idea on one topic introduces a value of ``one" into the square root function. Suppose we want to find the diversity of a set of three items. If all items in this set are on compost (\ie, a single cluster), the fitness will be $ \sqrt{1+1+1}=\sqrt{3}$, if we have two items covering compost and one on food festivals, the fitness will be $ 1+ \sqrt{2}$, while if all items cover different topics we will achieve the maximum diversity of magnitude $3$.
Hence, diverse sets are rewarded by this additive sub-modular function. 
In Eq.~\ref{eq_div}, the value $\sum_{i\in P_k}L_{i,j}$ implies that items more similar to other items in their cluster (representative items) receive higher reward when added to an empty set. This concept is similar to \cite{boim2011diversification} used in recommender system, which identifies a set of representative items, one for each cluster.

In general, finding the set of ideas that maximizes Eq.~\ref{eq_div} is difficult. In fact, it is NP-Hard since it is essentially a combinatorial optimization problem where the value of adding an idea depends on what other ideas one has already added. When solving such problems, a well-known limit due to Feige~\cite{feige2011maximizing} is that any polynomial-time algorithm can only approximate the solution to Eq.~\ref{eq_div} up to $1-\frac{1}{e}\approx 67$\% of the optimal.  However, this is where choosing a sub-modular function for Eq.~\ref{eq_div} comes in handy. It turns out that greedily maximizing a sub-modular function\textemdash \ie, selecting ideas one at a time such that each choice maximizes Eq.~\ref{eq_div} as much as possible\textemdash is guaranteed to achieve that $1-\frac{1}{e}$ bound. This makes greedy maximization of Eq.~\ref{eq_div} the best possible polynomial-time approximation to an otherwise NP-Hard problem. Equation~\ref{eq_div} uses this property to obtain strong results, and we also leverage similar properties of sub-modular functions later during ranking to create greedy rankings, as well as to improve the convergence of a global optimizer.

A key limitation of using clusters in Equation~\ref{eq_div} is that we need to know or estimate, which idea belongs to which cluster. In general, we will not know cluster assignments ahead of time and may need to estimate them using different clustering algorithms like K-means \cite{manning1999foundations}, Spectral Clustering \cite{ng2002spectral}, Affinity Propagation (AP) or domain knowledge. However, as we show in our supplemental material, the performance of Eq.~\ref{eq_div} drastically depends on both the number and accuracy of any clusters. Moreover, ideas may not fall neatly into mutually exclusive buckets. These limitations led us to consider the next approach which does not require explicit clustering but rather considers coverage as a kind of volume measurement over a continuous space.

\subsection{DPP-based Diversification}
\label{sec:dpp}

Determinantal Point Processes (DPPs), which arise in quantum physics, are probabilistic models that model the likelihood of selecting a subset of diverse items as the determinant of a kernel matrix. 
The intuition behind DPPs is that the determinant of $L_S$ roughly corresponds to the volume spanned by the vectors representing the items in $V$.
Points that ``cover'' the space well should capture a larger volume of the overall space, and thus have a higher probability. 
Viewed as joint distributions over the binary variables corresponding to item selection, DPPs essentially capture negative correlations.
They have recently been used \cite{kulesza2012determinantal} for set selection problems in machine learning tasks like diverse pose detection and information retrieval \cite{kulesza2011learning}.

While conceptually simple and fairly straightforward to compute, DPPs suffer from a couple of subtle numerical and optimization issues when used to rank-order items. We review and solve these in Sec.~\ref{sec:ranking_diversity}, but, briefly, the problems have to do with the sub-modularity and magnitude of the determinant when comparing growing set sizes.
Similar to sub-modular functions, one of the main applications of DPP is extractive document summarization, where it provided state-of-art results. 
As shown by Kulesza \etal \cite{kulesza2011k}, one of DPPs advantages is that computing marginals, certain conditional probabilities, and sampling can all be done exactly in polynomial time.

For the purposes of modeling real data, the most relevant construction of DPPs is through L-ensembles \cite{borodin2009determinantal}.
An L-ensemble defines a DPP via a positive semi-definite matrix $L$ indexed by the elements of a subset $S$.
The probability of a set $S$ occurring under a DPP is calculated as:
\begin{equation}
Div_2(S) = \frac{det(L_S)}{det(L+I)}
\label{eq:dppk}
\end{equation}

$L_S \equiv [L_{ij}]_{ij \in S}$ denotes the restriction of $L$ to the entries indexed by elements of $S$ and $I$ is $N \times N$ identity matrix.
For any set size, the most diverse subset under a DPP will have maximum likelihood $Div_2(S)$ or equivalently the highest determinant (the denominator can be ignored for maximizing diversity of a fixed set size). 
As the similarity between two items increases, the probabilities of sets containing both of them decrease.
Unlike the previous sub-modular clustering, DPPs only require the similarity kernel matrix $L$ and do not explicitly need clusters to model diversity. This also makes them more flexible, since we only need to provide a valid similarity kernel (\eg, image or shape kernels), rather than an underlying Euclidean space or clusters.

So what does this all mean for a designer?
Let us get back to our example earlier in the paper. If we represent the four ideas as TF-IDF vectors and compute their cosine similarity, we find that first two ideas related to compost have cosine similarity with each other of $0.61$. The similarity between other pairs of ideas is close to zero ($<0.1$). This is expected, as the first two ideas are based on compost and have little in common with other ideas that are based on food festivals and online web communities. When we compute the determinant of the sub-matrix for the first two ideas (numerator in Eq.~\ref{eq:dppk}), it is $\approx 0.62$, whereas for determinant of first and third idea is $\approx 1$. Hence, DPPs (via the numerator in Eq.~\ref{eq:dppk}) penalize set that contain similar ideas, without requiring us to define any explicit notion of a cluster. This flexibility (plus the strong comparative empirical performance we note in our supplemental material) is why we will use DPPs for our ranking algorithms and experiments in the rest of the paper.

\section{Ranking items}
\label{sec:ranking}

Thus far, we have compared and analyzed diversity metrics for sets of fixed size. In such cases, a diversity metric like DPPs will give the same value for any permutation of a set since it does not care about the order of the items within the set.
This is not desirable for rankings, where users browse sequentially through an ordered list of items up until they reach some (unknown) user-specific limit.
This section addresses how to adapt diversity and quality metrics to such cases and compute objective functions over ranked lists (or, equivalently, permutations over items in list). To the best of our knowledge, this is the first time DPPs have been extended to such cases, and doing so involves tackling some subtle but important properties of DPPs over growing set sizes.

\subsection{Extending DPPs to rank diversity on ordered sets}
\label{sec:ranking_diversity}

To extend DPPs to ranked lists, we first need to review some of the geometric intuition behind how the determinant calculations central to DPPs change as we grow the set size. Specifically, we need to look at the determinant of $L_S$, which is the portion of the similarity kernel ($L$) formed by the selected items ($S$). This square matrix grows as we add items to $S$. Mathematically, its determinant is the product of the eigenvalues of $L_S$. Geometrically, the magnitude of the determinant is the volume of the $|S|$-dimensional parallelepiped formed by the elements in set $S$. This implies that adding elements to a set decreases the determinant. 

This behavior creates two problems for ranking.
First, as we add items to a ranking, the determinants and thus our diversity measure do not have similar length-scales. This means we cannot directly compare or optimize rankings of different length, which matters if we wish to assemble ranked lists in a greedy fashion by progressively adding elements.

To circumvent this problem, we re-define diversity from Eq.~\ref{eq:dppk} to Eq.~\ref{eq:dppn} below:
\begin{equation}
Div_3(S) = {(det(L_S))}^{\frac{1}{n}}
\label{eq:dppn}
\end{equation}

This essentially scales the diversity of a set of size $|S|=n$ by its size.
Geometrically, $Div_3(S)$ is proportional to the side length of a $n$-dimensional cube with same volume as the parallelepiped.
For a given set-size, n is constant, so maximizing $Div_3(S)$ is equivalent to maximizing $Div_2(S)$. However, mathematically, $Div_3(S)$ is the geometric mean of the eigenvalues of $L_S$. It represents the central tendency or typical value of the set of eigenvalues via their product. 

A second problem with the determinant is that adding the same item to a short list versus a long list can create two issues: (1) Taking the sum of $Div_2(S)$ for a ranked list would not be accurate as items at the beginning of the list will have much larger impact on diversity compared to items down the list. (2) If two almost identical items are placed in the same set, then the determinant quickly collapses to zero (or close to it), introducing numerical errors that make it difficult to compare good versus bad sets on a finite-precision computer. To address this, we use the log-average to measure list fitness for sets of increasing size:
\begin{equation}
\begin{aligned}
    Div_R = \sum_{k=1}^N{\frac{log(det(L_{S(k)}))}{k}}\\
    L_{S(k)} \equiv [L_{ij}]_{ij \in [1,2,..k]}
\label{eq:dpprank}
\end{aligned}
\end{equation}
The monotonic nature of logs does not change the optimal set, but helps eliminate numerical and discounting errors during the computation of the diversity score.

Despite those computational issues, the determinant's behavior does have a useful side-effect. Because the determinant begins to collapse once the sets start to cover the space (\ie, additional vectors begin to lie close-by to existing vectors), it creates a natural diminishing marginal utility condition where, once we add sufficiently diverse elements, the rankings of further items are not as strongly influenced by item diversity. What this means is that, at some distribution-dependent point in the ranking, items further down the list can be sorted by quality only, with little to no change in the diversity score for the total ranking. This has substantial computational benefit because while computing diverse sets is NP-Hard and thus needs to be approximated, at a certain point we can switch over to a much simpler and optimal sorting task to produce the remainder of the ranking.

\subsection{Ranking Quality}
\label{sec:ranking_quality}
The recommended list of items should not only be diverse, but also of high-quality. High quality items ensure that they are relevant to the design problem. 
While finding the best quality metric for a set of items is still an active area of research, researchers have developed many tractable solutions, including crowd-voting \cite{toubia2007adaptive}, expert opinion \cite{mollick2015wisdom} or similarity to prior high-quality ideas \cite{ahmed:cscw_2016_winning}. 
Unlike diversity, evaluations of quality are independent, easy to parallel-process, and not combinatorial in nature; this makes estimating quality (comparatively) tractable using existing techniques. We assume that a quality rating is available for every item, or can be estimated (\eg, using our prior work on quality estimation \cite{ahmed:cscw_2016_winning}).

Given a quality rating for every item, we need to define the overall quality fitness for a ranked list. For this purpose, we use normalized discounted cumulative gain (nDCG) a standard ranking metric for relevance judgments in ordered lists \cite{jarvelin2002cumulated}.
It varies from $0$ to $1$, with $1$ representing the ideal ranking sorted by relevance. This metric is commonly used in information retrieval to evaluate the performance of ranked lists by giving more weight to results appearing at the top of list. If $k$ is the maximum number of entities that can be recommended, then $DCG_k$ is given by:
\begin{equation}
DCG_k = \sum_{i=1}^{k} \frac{2^{rel_{i}}-1}{log_2(i+1)}
\label{eq:dcgk}
\end{equation}

Here $rel_{i}$ is the relevance of $i^{th}$ item on the list.
$IDCG_k$ is defined as the maximum possible (ideal) DCG for a given set of items \ie, when items are sorted by relevance. Hence normalized DCG is given by:

\begin{equation}
nDCG_k = \frac{DCG_k}{IDCG_k}
\label{eq:ndcge}
\end{equation}
  
To get an intuitive understanding of $nDCG_k$, consider the following example. Assume that a challenge has 5 items and that we get two lists of 5 items each. Let the relevance ratings be [11, 5, 3, 2, 1] for these items respectively. We normalize these ratings to [1, 0.4, 0.2, 0.1, 0]. Now let us say that List 1 is represented as $[1, 2, 3, 5, 4]$ and List 2 is $[4, 1, 2, 3, 5]$. Using Equation~\ref{eq:ndcge}, $DCG_{5}$ for List 1 equals $1.304$ and $DCG_{5}$ for List 2 equals $0.927$. Here, an ideal list will be one where all items are sorted by the quality and $IDCG_{5}$ is 1.307. Hence, $nDCG_{5}$ for List 1 is $0.998$ while for List 2 is $0.709$. Using this metric, List 1 will be a preferred method as it provides more relevant (higher quality) items early on. Hence, we use $nDCG_N(r)$ as our measure of quality for different permutations $r$ of $N$ items.

\section{Optimization}
Now that we have ways of comparing the diversity and quality of different ranked lists, our task is to find the `best' ranking (equivalently, permutation) that trades off diversity and quality. One na\"{i}ve approach is to equally weigh diversity and quality, and then optimize over the joint objective. However, such an approach is too restrictive since a designer may prefer a ranking that encourages quality more than diversity, or vice versa. Also, in one domain, it is possible that the highest quality ideas are also the most diverse while in another domain, it may happen that one can achieve significant diversity gains by losing almost no overall quality.

It is difficult to unilaterally predict, for every domain, the appropriate trade-off between quality and diversity. Instead we approach ranking as a multi-objective optimization where we generate a entire trade-off front of different rankings\textemdash from purely maximum quality rankings to maximally diverse rankings\textemdash that allows a designer to choose the extent to which he or she wishes to encourage diversity over quality or compute how much overall quality (if any) he or she might sacrifice to encourage diversity (our below results suggest that such sacrifices are small). 

Multi-objective optimization is used widely where optimal decisions need to be taken in the presence of trade-offs between two or more conflicting objectives.
Without additional subjective preference information, all trade-off solutions are considered equally good.
Obtaining the trade-off front gives choice to a designer. For example, a designer may choose a highly diverse ranking during early-stage ideation to explore the design space and then later transition to rankings that more heavily weigh quality. Likewise, if a designer wants to ensure a minimum quality threshold among all obtained ranked lists, our approach allows such constraints. As far as we know, our single proposed ranking algorithm is the first to permit such flexibility when comparing and ranking ideas.

At first glance, getting even close to the optimal ranking seems daunting, if not impossible. Not only is the general optimization problem NP-Hard, but the fact that we have two objectives (diversity and quality) implies that we need to generate not one, but an entire trade-off-front, of solutions. Mathematically, we know that we will have to approximate the optimal solution to this combinatorial problem (if we want to compute it in polynomial time). To do this approximation, we employ a stochastic global optimizer that relaxes the combinatorial problem into a search over real-valued scores. By themselves, such optimizers do not perform well on permutation problems such as ranking; however, due to our careful choice of our diversity scores above, we are able to leverage the properties of sub-modular functions to construct a greedy algorithm that efficiently computes diverse rankings. This substantially accelerates convergence of the global trade-off-front.

\subsection{Single Objective Greedy Optimization}
\label{sec:greedy}
A ranking optimized for quality can be easily obtained by sorting ideas by quality. Hence, below we explore the more technically challenging task of ranking ideas for maximal diversity.
Many diversification methods like Maximum Marginal Relevance~\cite{carbonell1998use} use greedy search to obtain a ranked list of diverse items.
Likewise, we propose below a greedy algorithm for DPP-based diversity to find a diverse list of items. 

\begin{enumerate}
\item  \qquad $A = \emptyset $
    
\item \qquad $A = A \cup \{S_i,S_j\}$ s.t. $[i,j]=\argmin(L)$

 \item  \qquad {\bf while} ($U \neq \emptyset $) {\bf do} 

 \item  \qquad\qquad Pick an item $S_i$ that minimizes $det(L_{A \cup i})$

 \item  \qquad\qquad $A = A \cup \{S_i\}$

 \item  \qquad\qquad $U = U - S_i$

 \item  \qquad output $A$
 \label{alg:greedy}
\end{enumerate}

Here, the method greedily adds members to the set by maximizing the probability given by Equation~\ref{eq:dpprank}. Suppose $U = \{ 1,2,3,..N \}$ is a set of all $N$ items and $L$ is the $N \times N$ similarity kernel matrix. We find a diverse solution by greedily adding items to the empty set to maximize diversity of the obtained sets of increasing cardinality. As the logarithm of the determinant is sub-modular and monotonic, 
this greedy algorithm is theoretically guaranteed to provide the best possible polynomial time approximation to the optimal solution.
Our experimental results below also demonstrate that this greedy approach to DPPs leads to a higher diversity ranking compared to any random sample and even MMR.

\subsection{Multi-objective Global Optimization}
\label{sec:global_opt}

To optimize a permutation of a set of items, we use $N$ continuous variables mapped to a ranked list where each continuous variable $0 \le x_i \ge 1, i \in N$ is bounded. The permutation is obtained by sorting the variables. To understand the representation, consider the example below.
Let us assume that we have a set of $5$ items $V = {v_1, . . . , v_5}$.
Two possible candidate item score vectors might be $x_1$=[0.1, 0.3, 0.9, 0.5, 0.8] and $x_2$=[0.8, 0.2, 0.1, 0.4, 0.0]. 
On sorting by value, the corresponding ranks for $x_1$ and $x_2$ are $r(x_1)=[v_1, v_2, v_5, v_3, v_4]$ and $r(x_2)=[v_5, v_3, v_2, v_4, v_1]$, respectively. 
By changing the values of $x_i$, we can obtain any permutation of items. Note that the permutations are not unique and many $x_i$'s can map to the same permutation.

An ideal set of items should balance diversity and quality.
In a classical optimization approach, we could maximize any one of these two objectives directly by finding the best combination of items to recommend, subject to a given metric.
For both, however, we need to optimize across multiple, conflicting objectives. This involves finding sets of solutions that represent optimal trade-off between diversity and quality. We can then use those trade-off solutions to help designers explore and filter possible items.

In practice, one can use any multi-objective optimizer to explore those trade-offs. We chose to use Multi-Objective Evolutionary Algorithms (MOEAs), specifically the NSGA-II algorithm \cite{deb2002fast}. We generate the initial population randomly with a real valued gene of length $N$.
The real value indicates the rank relative to other items in the set.
The optimizer selects the next generation of the population using a solution's non-dominated rank and distance to the current generation to avoid crowding. Specifically, we use a controlled elitist genetic algorithm \cite{deb2002fast} with tournament selection, uniform mutation, and crossover.


\section{Results on Real-World Idea Data}
\label{sec:results}

We now demonstrate how the above methods can produce rankings for real-world design ideas. Specifically, we tested the proposed ranking on idea submission from OpenIDEO, an online design community where members design products, services, and experiences to solve broad social problems \cite{fuge:openideo_JCISE_2014}. We first describe the dataset and then demonstrate how to use our ranking method to produce idea lists that blend quality and diversity.

\subsection{Dataset}
On OpenIDEO, each challenge has a problem description and stages\textemdash \eg, Inspiration, Concepting, Applause, Refinement, Evaluation, Winning Concepts and Realisation\textemdash where the community refines and selects a small subset of winning ideas, many of which get implemented or funded. During the `Concepting' stage, participants generate and view hundreds to thousands of design ideas; in practice, the number of submissions make exhaustive review (even of the titles) impossible\textemdash \eg, for a medium-sized challenge of $\approx 600$ ideas, it would take a person over 25 hours to read all entries.\footnote{Assuming 200 words per minute at 60\% comprehension with the average OpenIDEO idea length of 500 words. This is conservative since many submissions also include images or videos.} 

To demonstrate our multi-objective optimization results on a concrete example, we use a challenge from OpenIDEO entitled \textit{`How might we better connect food production and consumption?'} The Food production challenge had total $606$ ideas with a vocabulary size of $1,656$ words and total $88,813$ words after pre-processing. 
For pre-processing the text data, we use standard natural language processing techniques to convert text to normalized word-frequency vectors (called TF-IDF vectors\cite{dong2005latent}). Specifically, we use a bag-of-words model to represent items as TF-IDF vectors. For pre-processing, we use Porter stemmer, Wordnet lemmatizer and remove stop-words. All words with inter-document frequency less than 1$\%$ and greater than $90\%$ are ignored.
We define the similarity between vectors ($L_{i,j}$) by computing the cosine-similarity between the TF-IDF vectors to get the similarity kernel $L$ or any sub-kernel $L_S$ for any subset of ideas $S \subseteq V$.

For any given idea, OpenIDEO has multiple metrics that indicate the quality of an idea: 
1) Applause\textemdash users can endorse an idea by pressing the `Applaud' button; 2) Citation count\textemdash users can cite ideas that inspired them, similarly to academic papers; 3) Comment or View count\textemdash each idea tracks the number of comments or views it receives; and 4)~a small set of winners proceed to the next stages and win the challenge\textemdash those that advance should correlate positively with quality. 
We use applause as our measure of quality since OpenIDEO uses applause as their own quality measure during Concepting stage. Applause of any idea i ($app_i$) is similar to Facebook `Like' feature, where community members endorse an idea.
We did not combine applause with views and comment count metrics as there is no straightforward way to determine optimum weights for combining these metrics. For example, it is difficult to argue if receiving more comments is more important as receiving more views. Secondly, we found that Applause had a Pearson's linear correlation of $0.65$ with views and $0.69$ with comment count, so choosing a different quality measure does not substantially alter our results. We evaluate our methodology using relevance defined in Equation~\ref{eq:dcgw}.
\begin{equation}
    rel_{i}= \frac{app_i - min(app)}{max(app) - min(app)}
    \label{eq:dcgw}
  \end{equation}

\subsection{Results}

For $606$ ideas, the number of possible permutations (\ie, rankings) is $606!\approx 10^{1424}$, which is impossible to compute exhaustively to obtain the ideal trade-off front.
We use NSGA-II for bi-objective optimization to simultaneously maximize DCG Applause defined in Eq.~\ref{eq:ndcge} and Diversity defined in Eq.~\ref{eq:dpprank}. 

\begin{figure}[t]
\begin{center}
\centerline{\psfig{figure=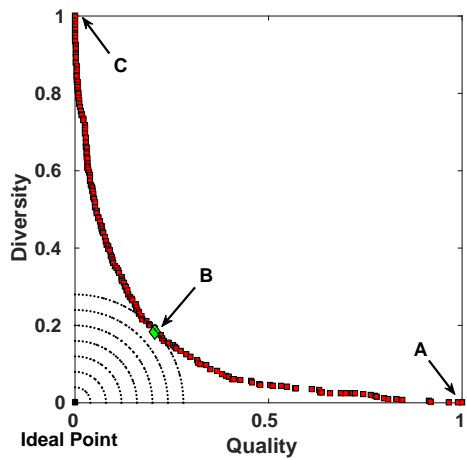,width=3.34in}}
\end{center}
\caption{Trade-off front between diversity and quality of ranked lists. Each point is a different permutation of 606 ideas.
A is the most diverse solution while C is the solution with highest quality objective. Indifference curves are used to find the Point B closest to the Ideal Point.}
\label{fig:radial}
\end{figure}

We use a population size of $500$ and run the optimization for $1000$ generations with crossover rate of $0.8$ and mutation rate of $0.01$.
Greedy solutions for applause and diversity are introduced into the population at first generation to speed up convergence.
We get $175$ unique points on the trade-off front.
The trade-off front between quality and diversity is shown in Fig.~\ref{fig:radial}. 
The values for both objectives are scaled between $0$ to $1$, with the optimization problem posed as minimization of both objectives.
Note that each point on the trade-off front is a permutation of all ideas\textemdash that is, each point on the trade-off front represents a different possible ranking (\ie, permutation) of the 606 ideas.

While this trade-off front lets a designer choose different rankings, depending on how much they prefer quality over diversity or vice versa, some designers may want just one ranking of ideas. To achieve this, we propose using indifference curves \cite{chiu2008hyper} for selecting an intermediate solution B on the trade-off front. After we normalize the objectives, every circle that uses the origin (\ie, the Utopia or Ideal point) as its circle center can be considered to be a true indifference curve. The points on smaller radius indifference curves are more desirable than those on bigger radius indifference curves. Therefore, the best solution is the point on the frontier that is tangent to the smallest valued indifference curve. In this way, indifference curves essentially weigh diversity and quality equally to provide a single ranking\textemdash point B. However, our approach can be easily adapted to different ratios of preferences by altering shape of the radial curves or even running a one-dimensional search along the trade-off front using techniques like interleaved comparisons \cite{hofmann2013fidelity} or knee region detection\cite{deb2011understanding}.

To compare the types of rankings produced by our proposed approach on a concrete example, let us take three points on the trade-off front marked as A, B and C. The maximum quality permutation C sorts ideas by applause while the maximum diversity permutation A is the one obtained by our above greedy search. We list the top 10 ideas in List A, B and C in Table~\ref{tab:abc}.
One can notice that solution C (ranked purely by highest applause) has no overlap with most diverse solution A. Reading through the ideas in A (the most diverse ranking), one can notice that despite being diverse, they are poorly written and somewhat irrelevant to the challenge. 
For example, idea titled ``Branded Clothing'' proposes referencing local producers on hats and t-shirts. It is a two line idea, without any details on implementation, practicality etc.
We found that these ideas often have poor quality scores as they did not address the challenge requirements, were not well written, and did not engage with the community in improving these ideas. Although permutation A is most diverse, suggesting such a set may not be useful for inspiring a designer. In contrast, the highest quality permutation C has several redundant ideas. The top 10 ideas in C have two similar ideas on mobile applications and multiple similar ideas related to farms. 
Our selected permutation B, by comparison, incorporates diversity by retaining seven high quality ideas from the most applauded set (C) and introducing three, one of which discusses schools adopting a program to source local food, another one of replacing fences with planted fruit trees, and a third one proposes traveling movie theater with local food. Having such a balanced list of high quality diverse ideas may be used to provide inspiration to designers to come up with designs.

\section{Discussion}
\label{sec:discussion}
Our ranking approach leads to two interesting observations: (1) A small selection of ideas is persistent along the trade-off front, and (2) studying the determinants of lists provides several insights into the nature of diversity and how diverse rankings compare to alternative rankings like highest-quality, MMR, or random permutations.

\subsection{Some ideas persist}
One key observation is that a small set of ideas persist in the Top-10 ranked items across the trade-off front. 
Taking the top 10 highest ranked items on all 175 lists obtained on our trade-off front, we find that they contain only 36 unique ideas as shown in Fig.~\ref{fig:top10}.
The titles of these ideas are reported in the supplement material.
This means that a designer can read only $6\%$ of the 606 ideas in the challenge, and still get a snapshot of ideas ranging from highest quality to most diverse. This also aligns with our previous observation in \cite{ahmed:idetc_2016_idea}, where a small subset of ideas were found to persist on the trade-off front for a different design problem. It is also interesting to note the ideas with very high frequency on the trade-off front like ``The Farmer and The Chef''.
The idea is both unique and high quality, due to which it is present in Top 10 ideas for $97\%$ of the lists on trade-off front.
One of this paper's ancillary outcomes is to identify such high quality unique ideas.

\begin{figure}[t]
\begin{center}
\centerline{\psfig{figure=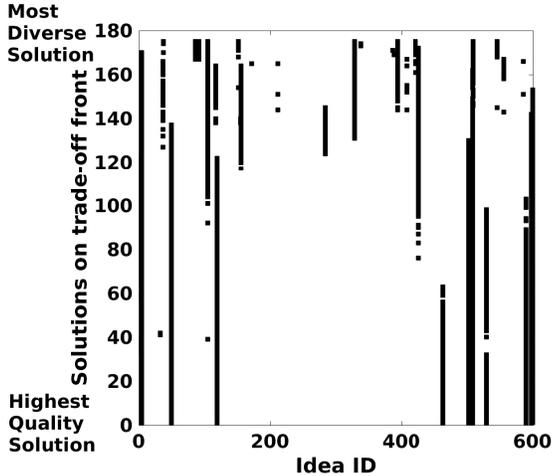,width=3.34in}}
\end{center}
\caption{Ideas selected in Top 10 of different solution sets on the trade-off front between quality and diversity. The figure shows that only a small set of 36 unique ideas appear on trade-off front (the lines in the figures).  On the bottom are ideas selected for high quality in the trade-off front, while top of the figure has ideas with high diversity}
\label{fig:top10}
\end{figure}

\begin{figure}[t]
\begin{center}
\centerline{\psfig{figure=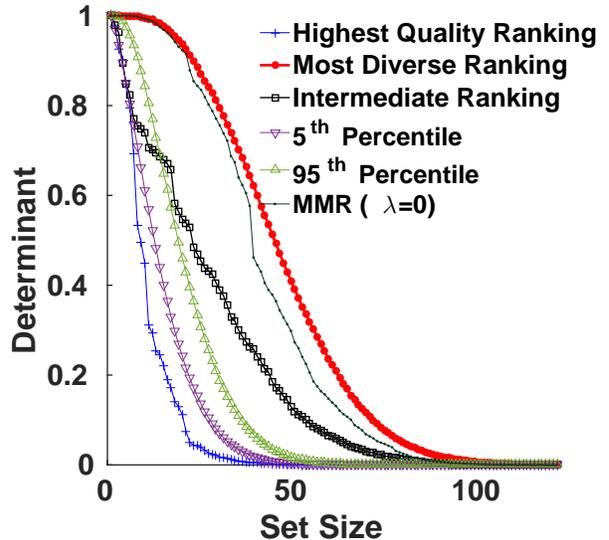,width=3.34in}}
\end{center}
\caption{Determinant of subsets for different ranked lists. The $5^{th}$ and $95^{th}$ percentile solutions show that marginal gain in diversity after 60 solutions is very low. The most diverse solution (A) from trade-off front selected using greedy solution is significantly more diverse than random permutations}
\label{fig:diverse}
\end{figure}


\subsection{Diversity matters less for larger sets}
Figure~\ref{fig:diverse} shows the determinants for ordered subsets of different permutations. That is, it plots $det(L_{S(k)})$, where as defined before, $L_{S(k)} \equiv [L_{ij}]_{ij \in [1,2,..k]}$, or how the determinant changes as you add ideas from progressively further down the ranked list. It includes the highest quality ranking (C), the most diverse ranking (A), and our intermediate ranking (B). 
To compare our greedy algorithm with existing methods in the literature, we also plot the maximum diversified permutation
using MMR \cite{carbonell1998use} with $\lambda=0$, as well as $5^{th}$ and $95^{th}$ percentile from 5000 random permutations to compare to random chance.
Figure~\ref{fig:diverse} provides four insights into using determinants as diversity metric.

First, Fig.~\ref{fig:diverse} shows that our diverse greedy list outperforms both randomized rankings and MMR, in terms of promoting diverse rankings. 

Second, We can see that the most applauded set is below the $5^{th}$ percentile of diverse sets. This shows that, for this challenge, ranking ideas purely by quality produces a ranking that lacks diversity, even compared to random rankings. On other hand, using the greedy solution to obtain solution A (or even our intermediate solution B) leads to big gains in diversity, significantly even above the $95^{th}$ percentile. This indicates that our greedy algorithm is efficiently finding a diverse solution. 

Third, the determinants collapse to zero for at most 100 items in the ranked list. This implies that there is not much marginal gain in diversity once one has added many items (\ie, beyond 100)\textemdash this makes sense since, by that point, new items will not drastically change the geometric mean of the volume spanned by the determinant. This also allows us to save computational effort by only maximizing Eq.~\ref{eq:dpprank} up to $N=100$ and then sorting by quality further down the list. This exact $N$ cutoff will be problem dependent; however, Fig.~\ref{fig:diverse} is one  criterion for determining when that transition takes places.

Lastly, one can also notice that the determinant magnitude decreases as set size increases. This intuitively makes sense since Eq.~\ref{eq:dpprank} scales the diversity of sets of different sizes by using geometric mean. Thus, simple area under this curve will prioritize diversity in elements early on in the ranking.

\subsection{Limitations and Future Work}
We provided a tractable, computational ranking method that simultaneously maximizes a trade-off between quality and diversity of items. As a byproduct, this ranking can also produce diverse, high-quality subsets (such as top 10 lists). 
However, the method has a few limitations where more research focus is needed.

First, selecting the ``correct'' diversity kernel to identify similar items is key to the success of any diversification method. 
At a conceptual level, our main assumption is that the kernel that encodes what makes ideas similar or different is good or accurate.
We used a standard cosine similarity kernel for comparing text, however applying machine learning techniques to learn this kernel based on human perception of diversity may improve performance \cite{kulesza2011learning}. Also, this method is only able to compute the diversity of ideas within the set of the current data. If all global ideas are considered, the similarity kernel and clustering will change, which will affect the diversity metric evaluations.\footnote{To some extent, using humans to construct the diversity kernel may capture this global context, however one open research problem is determining when or for what types of problems that is true.}

Second, we assume that high quality items measured by crowd-voting is desirable for inspiring designers to come up with new designs. The rationale was that items which are more creative and better at addressing the design problem are voted
up by the crowd and are good candidates to inspire a designer. This assumption may become invalid if there are other latent factors affecting crowd-voting.
However, the main contributions of the paper are not really affected by choice of quality metric, since we assume a quality function (however one wants to define it) is available and the contributions are really how to do optimal ranking given such functions.

Third, but related to the second, is that we assume that we have quality estimates for all items. When this is not the case (\ie, the cold-start problem) we would need to approximate quality by content-based features like item uniqueness. For example, Ahmed \etal \cite{ahmed:idetc_2016_idea} showed that for OpenIDEO challenges, uniqueness of item and applause are strongly correlated and hence latter can be used in absence of former.

Lastly, our experiment only used text content to represent ideas. This representation was used to facilitate straightforward similarity computation and to demonstrate the key contributions of the paper.
In real cases, however, many ideas are a combination of text, images and videos, and only computing similarity using text may give an incomplete picture. 
The proposed method works for design ideas expressed in a variety of ways (text, sketches, function structure graphs, mixed-media, etc.) as all of the important contributions of our method\textemdash including how we calculate diversity, the sub-modularity conditions, our greedy approximation, the ranking algorithm, etc.\textemdash ultimately only depend on a similarity matrix between ideas (which we called $L$).
If one believes that humans might be the only reliable means to achieve some ground truth understanding of true idea diversity, then this is not a problem for our ranking method; simply use any existing metric- or kernel-learning algorithm to construct $L$ from human evaluators and then apply our ranking method to that new $L$.

Future research can focus on better methods to compute similarities. For example, one could compute metric spaces over visual designs \cite{chen2016designs, yumer2015procedural, burnap2016improving} and combine those with text similarity. In cases where it is difficult or undesirable to compute item features directly, one could use human judgments to compute item similarity (\eg, using techniques like ordinal embedding \cite{jain2016finite}) and directly substitute this similarity measure into Eq.~\ref{eq_div} above.


\subsection{Implications for Design Research}
Our proposed ranking method applies whenever a designer, team, or decision maker in an organization needs to sift through many ideas. This problem occurs in several design situations: 1) during ideation when multiple designers might generate many hundreds of possible ideas\textemdash be they text- or sketch-based ideas; 2) when large organizations wish to gather possible ideas or solutions from employees of their companies, for example via internal innovation tournaments~\cite{von2005democratizing}; 3) when companies or designs wish to solicit ideas from crowd-sourcing or online communities; and 4) when a designer wishes to use some kind of computational design synthesis system~\cite{chakrabarti2011computer} to generate thousands of possible solutions and then review the output such that he or she understands the scope or diversity of the solutions the system produces.
For those above situations, our paper has the following implications. 

First, 
our method is the first to enable polynomial time ranking of ideas by both quality and diversity with both provable performance guarantees and flexible control over how importantly the algorithm weighs diversity with respect to quality. Such capabilities matter when, for example, designers wish to promote diversity early on in a design process to enable divergent thinking, but then slowly move towards quality convergence over time. Our method provides an easy-to-understand parameter (namely the location along the trade-off front) that allows a designer to adjust how much they care about idea diversity.

Second, our approach provides a concrete metric (namely the difference in the determinant curves in Fig.~\ref{fig:diverse}) that allows a designer to assess the differences between the most-diverse and highest-quality rankings, and after how many ideas they have sufficiently covered the available space of ideas.
Such observations can provide useful knowledge about a given design problem domain. If our diversity metric plateaus very quickly, it indicates that the domain has very few unique topics. On the other hand, if it plateaus much later, the space of ideas likely has many different topics. Likewise, while not the focus of this paper, our method permits a new straightforward comparison of design exploration methods for a given problem; that is, given two methods, by comparing their curves in Fig.~\ref{fig:diverse} we can quantitatively study the extent to which different exploration methods cover wider portions of a design space. This allows us to gain new knowledge about both a given design domain as well as different processes designers use to explore it.

Lastly, while our paper only addressed trade-offs between quality and diversity, there is no technical reason why our proposed ranking algorithm and methodology could not also incorporate other useful design objectives\textemdash \eg, novelty, feasibility, \etc\textemdash provided such objectives can be evaluated efficiently on a large number of ideas (\eg, via expert or crowd ratings, or using computational evaluation where possible).
To enable practitioners deploy this method for their own domain, we have provided the source code \footnote{\url{https://github.com/IDEALLab/ranking_diversity_jmd_2017}} and encourage interested readers to use it. To get a trade-off front for any collection of design ideas, a practitioner needs only two inputs--- quality ratings for all ideas and a positive semi-definite similarity kernel, showing how similar ideas are to each other. However, the similarity kernel should be chosen carefully, as the diversity is evaluated on the same attributes for which similarity is calculated. For example, let us say a practitioner wants to apply our method to a collection of sketches. Suppose they use similarity kernel based on a surface feature like the color used to sketch the idea. In such a case, the diverse ranking will also output a ranked list, which has sketches of different colors at the top of the list. In contrast, if they use similarity based on some feature like the mechanism used, the ranked list will reflect the same attribute. 

\section{Conclusion}

In this paper, we proposed a method to measure diversity of sets and ranked lists of items. 
These measures were combined with quality to simultaneously maximize the quality and diversity of a ranking.
Specifically, the paper added the following new pieces of knowledge: 1) how to extend set-based diversity metrics to rank-based diversity measures, 2) how to rank ideas by diversity in polynomial time using a greedy strategy with theoretical performance guarantees, 3) how to trade-off quality and diversity when ranking ideas, and 4) how one can use the determinant of a design space to uncover properties of that space (such as how much quality one has to sacrifice to gain diversity) and the extent to which one can achieve compression in the ideas one considers (via comparisons along the quality-to-diversity trade-off front).

We demonstrated and validated the above contributions using both benchmark datasets and 606 real-world design ideas from an OpenIDEO challenge. We showed that our method produces higher quality, more diverse rankings than competing techniques. 
Our findings have several implications both for ranking items and studying ideation at large scale.

First, Fig. \ref{fig:top10} showed that, out of 606 ideas, only 36 unique solutions appeared across any portion of the trade-off front in Top 10 ideas, from high-quality to high-diversity. This implies that, even without picking a location on the trade-off front, we can achieve substantial compression in the ``minimal set" of inspiring ideas a designer might consider\textemdash roughly 6\% in our example. In the real-world scenario we analyzed, this meant reducing designer effort from roughly 25 hours to 90 minutes.

Second, when trading off diversity and quality, we found that maximizing diversity without considering quality produced less useful ideas than considering the combination. This implies that we need better automated quality metrics for ideas\textemdash similar to those researchers have proposed for diversity or variety\textemdash if we hope to scale up our ability to evaluate or inspire creative ideas. 

\newpage
\appendix       
\begin{table*}[t]
\caption{OpenIDEO ideas on trade-off front}
\footnotesize
\centering
\label{table_ASME}
\begin{tabular}{|p{7cm}|p{4cm}|}
\hline
Title & Set (Most Diverse (A), Highest Quality (C) and Radial Set (B))\\
\hline
Building `Transparency' App (updated) & C \\
Eatcyclopedia: A Phone App to Help Connect and Inform & C\\
Hold Seasonal ``Open House" Days at Local Farms & C\\
The Farmer and The Chef & C, B\\
Closing the Farmers Market Loop & C, B\\
Market Days + Food Trucks = Serving Low-income Neighborhoods & C, B\\
Redesign the supermarket layout based on food miles... UPDATED & C, B\\
Window to the Farm & C, B\\
Public Kitchen & C, B \\
A celebration of imperfection & C, B \\
 50 Within 50 & B\\
Traveling Movie Theater on Farms &  B\\
Fruit Trees instead of Fences & B, A\\
Branded Clothing & A\\
Intensive two-week Internship on farms: Interns will teach others when they come back to the city & A\\
Trick yourself into sustainable buying & A\\
Trade \& resell network for CSA share-holders. Specific to central pick-up location for many CSA programs. & A\\
Dentell & A\\
Install Greenhouses at Train Stations & A\\
A new youth movement: Healthy Eating and living & A\\
fruity roofs & A\\
Hack Cooking to Make it Appealing & A\\
\hline
\end{tabular}
\label{tab:abc}
\end{table*}

\bibliographystyle{asmems4}

\bibliography{asme2e}

 \appendix       
  \section{Application to Sketches}
 \label{sec:sketchapp}
 
\begin{figure}[!htb]
\begin{center}
\centerline{\psfig{figure=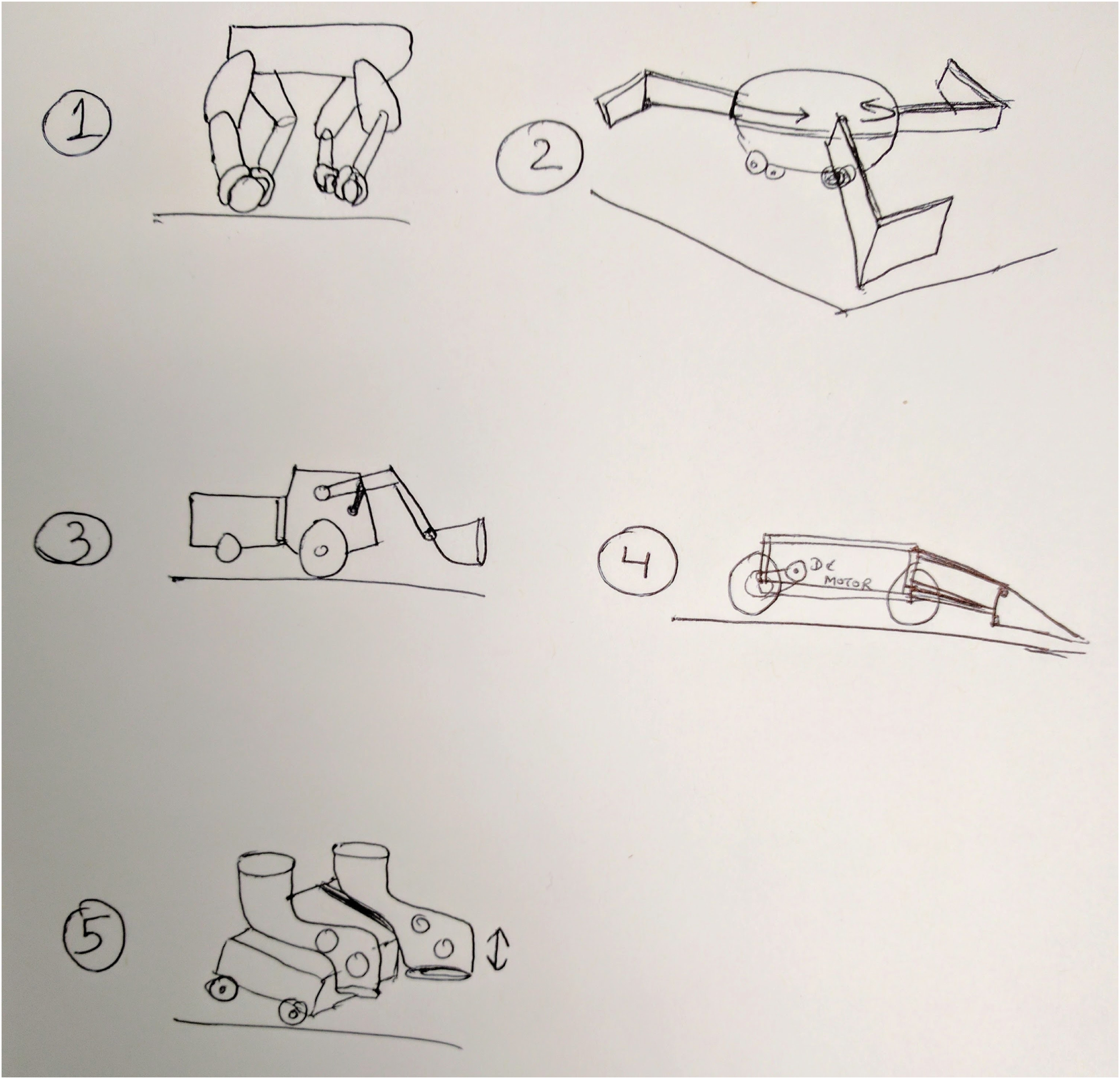,width=3.34in}}
\end{center}
\caption{Five sketches of semi-autonomous device to collect golf balls from a playing field.}
\label{fig:sketcheg}
\end{figure}

To demonstrate the applicability of our method to non-text design problems, we take a simple example of ranking five sketches.  
We adopt the design problem discussed in \cite{shah2003metrics}, where one has to sketch semi-autonomous device to collect golf balls from a playing field and bring them to a storage area. Inspired by the sketches in \cite{shah2003metrics}, five sketches for possible devices are sketched by one person, as shown in Fig.~\ref{fig:sketcheg}. The sketches are numbered $1$ to $5$.

To apply our method, we need the quality ratings and similarity kernel for these sketches. Unlike text ideas, these sketches are not represented as vectors. Hence, we solve a sub-problem of estimating the similarity between sketches using a human rater.

To do so, we decide to learn an embedding of
data based on similarity triplets of the form, ``Sketch A is more similar to Sketch B than to Sketch C''.
To find the similarity between these sketches, 
we ask a human rater to give his relative preferences as shown in Table~\ref{tab:triplet}.
The rater is asked to provide ten comparisons, where he specifies which sketch is closer to the base image. So rating provided in row 1 of Table~\ref{tab:triplet} implies that Sketch 3 is more similar to Sketch 1, compared to Sketch 2.
Using these triplet ratings, we learn two dimensional embedding for all sketches using t-Distributed Stochastic Triplet Embedding (t-STE) \cite{van2012stochastic}. The model is used to obtain a truthful embedding of the underlying data using human judgments on the similarity of objects. Essentially, the model takes as input the triplet embeddings shown in Table~\ref{tab:triplet} and generates a lower dimensional vector embedding for each sketch.

\begin{figure}[!htb]
\begin{center}
\centerline{\psfig{figure=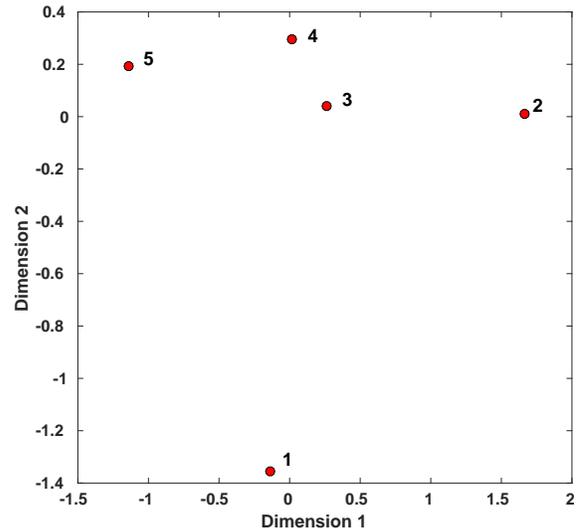,width=3.34in}}
\end{center}
\caption{Two dimensional embedding of five sketches calculated using t-Distributed Stochastic Triplet Embedding. It shows sketches 3 and 4 are similar to each other, while 1, 2 and 5 are unique.}
\label{fig:sketch2d}
\end{figure}

\begin{figure}[!htb]
\begin{center}
\centerline{\psfig{figure=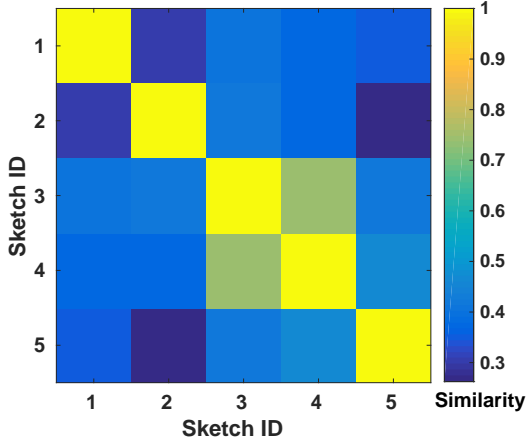,width=3.34in}}
\end{center}
\caption{Similarity kernel for five sketches calculated for 2-D embedding}
\label{fig:sketchsim}
\end{figure}

Fig.~\ref{fig:sketch2d} shows the output of t-STE model---a two dimensional embedding for the five sketches. From the embedding, one can conclude that Sketch 1 is quite unique (far away from all other sketches).
Using distances from this embedding, we calculate a similarity kernel shown in Fig.~\ref{fig:sketchsim}. 
From the similarity kernel and the two dimensional embedding, one can notice that the rater found Sketch 3 and 4 similar to each other, while sketch 1, 2 and 5 are relatively unique. Having obtained the positive semi-definite similarity kernel, next we find quality ratings for all the sketches.

We ask a human rater to provide quality ratings for the sketches on a scale of 1 to 10, with 10 being the highest quality idea. The quality rating provided by the rater for these sketches are 3, 2, 7, 8 and 6 respectively. Using these ratings, if we sort these sketches in descending order of quality, we obtain the following ranking: 4, 3, 5, 1 and 2.

\begin{figure}[!htb]
\begin{center}
\centerline{\psfig{figure=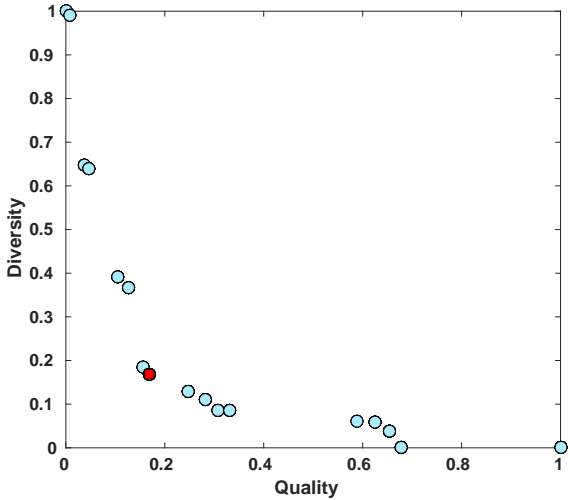,width=3.34in}}
\end{center}
\caption{Trade-off between Quality and Diversity for Ranking of five sketches}
\label{fig:sketchtradeoff}
\end{figure}

Using the quality ratings and similarity kernel as inputs to our method, we calculate the trade-off front between diversity and quality as shown in Fig.~\ref{fig:sketchtradeoff}.
There are $17$ unique solutions on the trade-off front. We also find the intermediate solution using indifference curves (shown using red marker).
Below are the highest quality, highest diversity and the intermediate rankings on trade-off front:

\begin{flushleft}
$\bullet$ Ranking by Quality: 4, 3, 5, 1, 2.\\
$\bullet$ Intermediate Ranking: 4, 5, 2, 1, 3.\\
$\bullet$ Ranking by Diversity: 2, 5, 1, 4, 3.
\end{flushleft}

From the rankings obtained, one can verify that ranking by quality (left extreme of trade-off front) has sketches sorted by quality ratings. For the most diverse ranking (right extreme of trade-off front), the method gives higher ranking to the unique sketches 2, 5 and 1, followed by similar sketches 4 and 3. Finally, the intermediate ranking balances quality with diversity.

While this example was simple and only 120 permutations were possible for a small set of five sketches, it demonstrated a straightforward way to adapt our method for a sketch based design problem by first estimating the quality and similarity and then generating the trade-off front.

\begin{table}[]
    \centering
    \begin{tabular}{|l|l|l|}
    \hline
     Sketch A & Sketch B  &  Sketch C\\
     \hline
     1 & 3 & 2\\
     1  &4  & 2\\
     1 & 5 & 2\\
     1 &3 & 4\\
     1 & 3 & 5\\
     1 & 4 & 5\\
     2 &3 & 4\\
     2 & 3 & 5\\
     2 &  4 &5\\
     3 &  4 & 5\\
     \hline
    \end{tabular}
    \caption{Triplet embedding ratings provided by human rater. For each row, item in Sketch A column is more similar to item in Sketch B column than Sketch C column}
    \label{tab:triplet}
\end{table}

\newpage
\section{Comparing Diversity Measures}
\label{sec:diversitycomparison}
To select the right diversity metric, we compare how accurately the DPP-based $Div_2(S)$ and sub-modular-function-based $Div_1(S)$ metrics capture diversity on a two-dimensional data set, where results can be verified by known ground-truth clusters. This helps us in discussing each method's advantages and disadvantages.

\subsection{Fixed Set Size Comparison}

\begin{figure}[t]
\begin{center}
\centerline{\psfig{figure=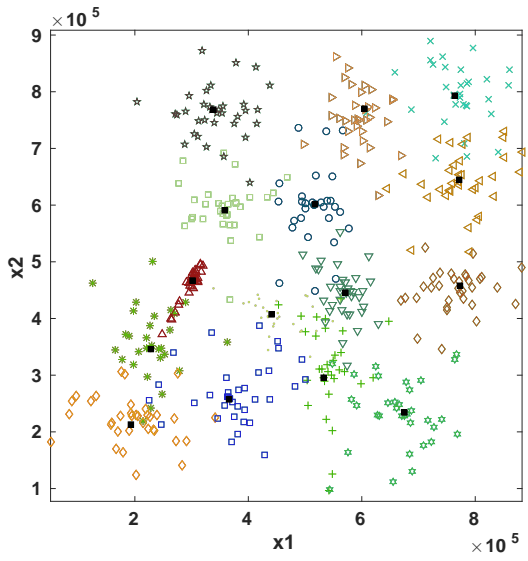,width=3.34in}}
\end{center}
\caption{Dataset with 500 points in 15 clusters}
\label{fig:s3}
\end{figure}

We use an existing clustering dataset shown in Fig.~\ref{fig:s3}.
It is a two-dimensional dataset with 500 data points across 15 clusters and has traditionally been used to compare clustering algorithms \cite{Ssets}. We use it to compare proposed diversity metrics under the criteria that a set is diverse if it has items from different clusters. This clustering interpretation is widely used in recommender systems for partitioning user profiles \cite{Zhang:2009:NIR:1632190.1632398} and information retrieval for grouping search intents \cite{chapelle2011intent}.
In Fig.~\ref{fig:s3} each point is allocated to a cluster and the cluster centers are plotted by black square markers. 

Suppose we want to select a diverse set of 8 points. Under our criterion, we would prefer to pick points from 8 different clusters; selecting multiple points from the same cluster would be less diverse. Mathematically, this cluster coverage can be quantified using Shannon entropy \cite{jost2006entropy}. Entropy measures the level of impurity in a group and will be maximum when each cluster has same number of elements and will be minimum if a single cluster has all the elements and other sets are empty. We considered a diversity metric `better' if it provides a higher fitness to a more entropic set (\ie, favors points from different clusters in our gold standard cluster datasets). To assess this, we created two sets of points, Set 1\textemdash high entropy, diverse, plotted using black squares\textemdash and Set 2\textemdash lower entropy, less diverse, plotted using red diamond markers. We then compare under what conditions the two methods agree that Set 1 is more diverse than Set 2.

\begin{figure}[t]
\begin{center}
\centerline{\psfig{figure=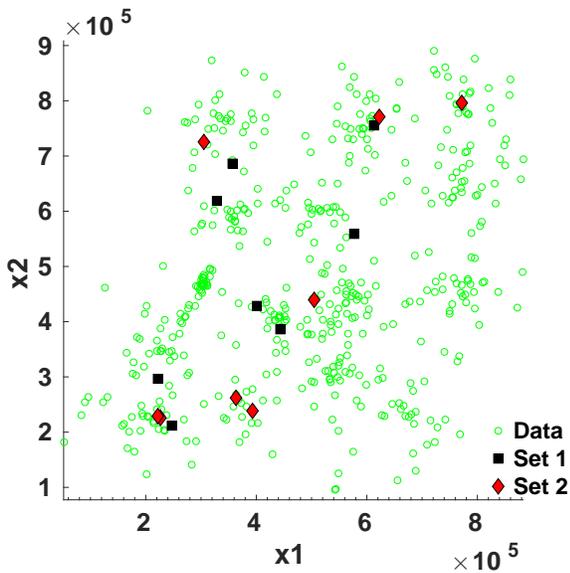,width=3.34in}}
\end{center}
\caption{Two sets of 8 points. 
Set 1 is more diverse than Set 2, as it has points in 7 clusters while 
Set 2 has points in 5 clusters}
\label{fig:set1}
\end{figure}

Figure~\ref{fig:set1} compares the above metrics by plotting two set of 8 points each.
Set 1 (the sub-modular clustering method) uses black square markers while Set 2 (DPPs) uses red diamond markers. Set 1 is more entropic than Set 2 it has 8 points belonging to 7 unique clusters while Set 2 has 8 points belonging to only 5 unique clusters.

For the DPP similarity measure between points we use a radial basis function (RBF) similarity kernel.
This similarity measure used gives score close to 1 to points which are nearby and low scores to distant points.
For Eq.~\ref{eq_div}, we need the similarity matrix and the cluster labels for each data point.
As a fair comparison, we use the same similarity kernel used for DPPs, but varied the clustering method and number of clusters since this method's performance depends on the clustering labels used for each data point. Specifically, we tested using the already known ground truth cluster labels (\ie, knowing the true clusters ahead of time), and the more realistic condition of computing the clusters using two methods: Spectral Clustering with $5$, $10$, $15$, or $20$ clusters, and Affinity Propagation (AP), which estimates the number of clusters from the data (it estimates $37$ clusters for this data set).

When we use the true $15$ Gold standard clusters provided with the data set, as expected, the measure agrees with Entropy, which is also defined using the same labels. 
When we use the similarity matrix defined before and apply Spectral clustering on it for $5$, $10$, $15$ and $20$ clusters, the results vary in agreement with entropy. Surprisingly, when the clustering is done with $15$ clusters but using Spectral Clustering instead of pre-known clusters, the method finds Set $2$ more diverse. We also use Affinity Propagation for clustering, which does not require pre-specifying the number of clusters and it finds $37$ clusters in the dataset.

For the DPP metric, we find that $det(L_{Set1})~>~det(L_{Set1})$, implying Set 1 more diverse than Set 2 as shown in Table~\ref{tab1}. This agrees with our entropy criterion. For sub-modular clustering, its performance was particularly sensitive to number of clusters used. When provided with the true cluster labels, as expected, it agrees with entropy. When it had to estimate the cluster labels, performance varied. Surprisingly, even when told to estimate the correct number of clusters (15), this particular choice of clustering algorithm negatively affected performance. It is possible that a different clustering algorithm (other than Spectral or AP) might offer more robust performance; our point here is that sub-modular clustering is particularly sensitive to how points are clustered and it is not immediately obvious how to verify one has made the ``right" choice on a problem with unknown ground truth. 

\subsection{Growing Set Size Comparison}
\begin{figure}[t]
\begin{center}
\centerline{\psfig{figure=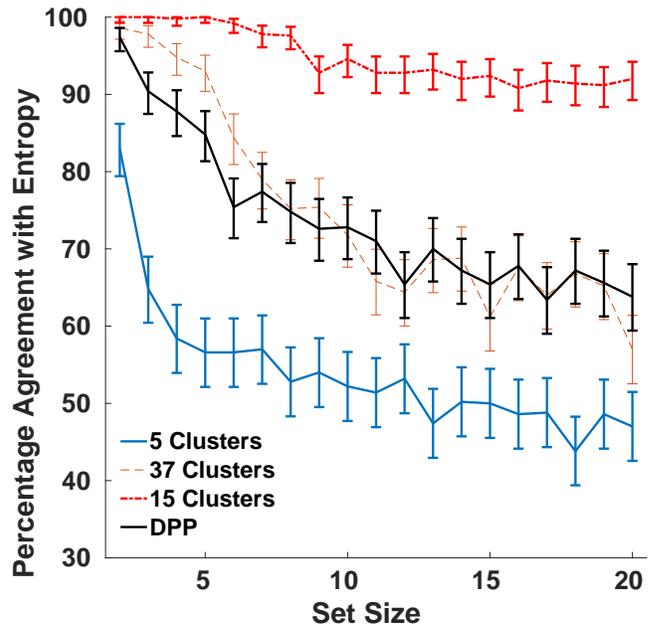,width=3.34in}}
\end{center}
\caption{Comparison of Sub-modular and DPP Diversity metrics for percentage agreement with Entropy. Random clusters of different sizes are used.}
\label{fig:clus}
\end{figure}

How does the above performance difference change if we change the size of the set?
Intuitively, if we are given two sets of two points each, it should be easier to estimate which is more diverse compared to when we have 20 points in each set. Figure~\ref{fig:clus} compares DPPs with sub-modular clustering methods as we vary the set size from $2$ to $20$. We randomly picked $1000$ sets of that size and divided those sets into two groups of $500$ each. We then conduct $500$ comparisons using one item from each group. We calculate the fitness using each method and record how often each methods agrees with entropy (our ground truth measure). Better metrics should agree with entropy more often and should consistently agree as the set size increases.
For clarity, we have shown four cases in Fig.~\ref{fig:clus}. For sub-modular clustering, using five clusters performs as poor as random chance, while using the known gold standard $15$ clusters obtains the best performance, as expected. 
The DPP diversity metric performs similar to Sub-modular diversity with $37$ clusters found using Affinity Propagation algorithm.

What do this results imply? Given the known clusters, sub-modular clustering has better agreement with our entropy success criterion than those based on DPPs. However, DPPs had more robust performance; that is, if we do not know the exact clusters ahead of time, DPPs perform better on average than sub-modular clustering.
In real world datasets, gold standard cluster labels are rarely available. Even estimating the number of clusters in a collection of design items is difficult. Hence, in such scenarios the parameter-less DPP method is a more robust choice for measuring diversity since using the incorrect number of clusters causes sub-modular-based metrics to perform poorly. However, if a good estimate of number and label assignments for clusters is available, then sub-modular clustering diversity performs well. In the paper, we use DPPs as our diversity metric since we assume that we do not know the number of clusters.

\begin{table}[]
    \centering
    \begin{tabular}{|l|l|l|}
    \hline
     Method & Set 1 Fitness  &  Set 2 Fitness\\
     \hline
     Unique Clusters & 7 & 5\\
     Entropy  &1.91  & 1.73\\
     DPP & 0.0611 & 1.8509e-04\\
     5 Clusters & 0.2201 & 0.2123\\
     10 Clusters & 0.2824 & 0.3043\\
     15 Clusters (Gold) & 0.3289 & 0.3043\\
     15 Clusters & 0.2989 & 0.3043\\
     20 Clusters & 0.3289 & 0.3043\\
     37 Clusters &  0.3289 & 0.3043\\
     \hline
    \end{tabular}
    \caption{Diversity fitness value using different metrics}
    \label{tab1}
\end{table}

\begin{table}[]
\begin{center}
\footnotesize
\begin{tabular}{|p{7cm}|}
\hline
Title of idea\\
\hline
(UPDATED) 'I am not from far away' label \\
The Farmer and The Chef\\
 Regional Food System+Commercial Kitchen+Food Entrepreneur Incubator=Collaborative Community\\
 Volunteer Farms Corp - like peace corp . . .or voluntary armed forces\\
Trick yourself into sustainable buying\\
Closing the Farmers Market Loop\\
Intensive two- week Internship on farms : Interns will teach others when they come back to the city\\
Dentell\\
Fruit Trees instead of Fences\\
 The treatment of a tomato\\
  Redesign the supermarket layout based on food miles... UPDATED\\
 Hack Cooking to Make it Appealing\\
  back to basics: bento recyclable trays to transport food instead of plastic bags.\\
 Traveling Movie Theater on Farms\\
 Roll-out Veg Mat\\
 Incentivizing Shifts from Lawn Service to Edible Gardening\\
Create Instant Farms on Vacant Lots\\
 Install Greenhouses at Train Stations\\
  Shopping list audit -- incentives for new shopping behavior\\
Carbon credit for local produce\\
 Make Veggie Topiaries\\
A new youth movement: Healthy Eating and living\\
  The Importance of villages\\
  fruity roofs\\
  50 Within 50\\
  Eatcyclopedia: A Phone App to Help Connect and Inform\\
  Market Days + Food Trucks = Serving Low-income Neighborhoods\\
  A celebration of imperfection\\
  Branded Clothing\\
  Hold Seasonal "Open House" Days at Local Farms\\
  Trade \& resell network for CSA share-holders. Specific to central pick-up location for many CSA programs.\\
  Zoning Bylaws To Permit Urban Beekeeping/Chickens\\
  iPhone, iPhone, what shall I cook tonight?\\
  Building 'Transparency' App (updated)\\
  Window to the Farm\\
  Public Kitchen\\
 \hline
\end{tabular}
\caption{All 36 ideas on trade-off front shown in Figure 2 in paper}
\end{center}
\end{table}

\end{document}